\documentclass[modern]{aastex61}
\usepackage{amsmath}
\usepackage{graphicx}
\usepackage{epsfig}
\usepackage{IEEEtrantools}
\usepackage[T1]{fontenc}
\usepackage{times}
\usepackage{lineno}
\newcommand\edh{\text{\dh}}

\begin{document}

\title{ Non-Axisymmetric Aberration Patterns from Wide-Field 
Telescopes using Spin-weighted Zernike Polynomials}

\correspondingauthor{Stephen Kent}
\email{skent@fnal.gov}

\author{Stephen M. Kent}
\affiliation{Fermi National Accelerator Laboratory \\
MS/127 \\
P.O. Box 500 \\
Batavia, IL 60510, USA}

\begin{abstract}

If the optical system of a telescope is perturbed from rotational symmetry,
the Zernike wavefront aberration coefficients describing that system
can be expressed
as a function of position in the focal plane using spin-weighted Zernike
polynomials.  Methodologies are presented to derive these polynomials
to arbitrary order.
This methodology is applied to aberration patterns produced by
a misaligned Ritchey-Chr\'etien telescope
and to distortion patterns at
the focal plane of the DESI optical corrector, where it is shown to provide
a more efficient description of distortion than conventional expansions.

\end{abstract}


\section{Introduction} \label{sec:intro}

\smallskip
Uncorrected aberrations in telescope optics often affect one's ability
to extract science from images at the telescope focal plane.
For example, in
weak lensing science, where one measures distortions in the images of galaxies
induced by gravitational lensing from foreground mass concentrations,
one's measurements are particularly
affected by low-order aberrations such as coma and astigmatism, which
induce ellipticities in images that mimic the effects of gravitational
lensing \citep[eg.,][]{jarvis08,jee11,hamana13}.  Furthermore, the operation of
certain instruments,
such as fiber positioners systems used for multiobject spectroscopy, require
detailed knowledge of the focal plane distortion pattern
\citep{akiyama08,kent16}.  It is often desirable to have an analytic
model to parametrize the aberration pattern across the focal plane.
For an axisymmetric system,
the aberrations depend only on radial distance from the symmetry axis, 
and thus the patterns can be characterized with 1$-$d radial polynomials.
However, even for well-aligned telescopes, the patterns will generally
have non-axisymmetric components, so a more general formulation is
desired.  This paper presents such a formulation based on the use
of spin-weighted functions.

\section{Theory} \label{sec:theory}

\begin{figure}
\centering
\epsfig{file=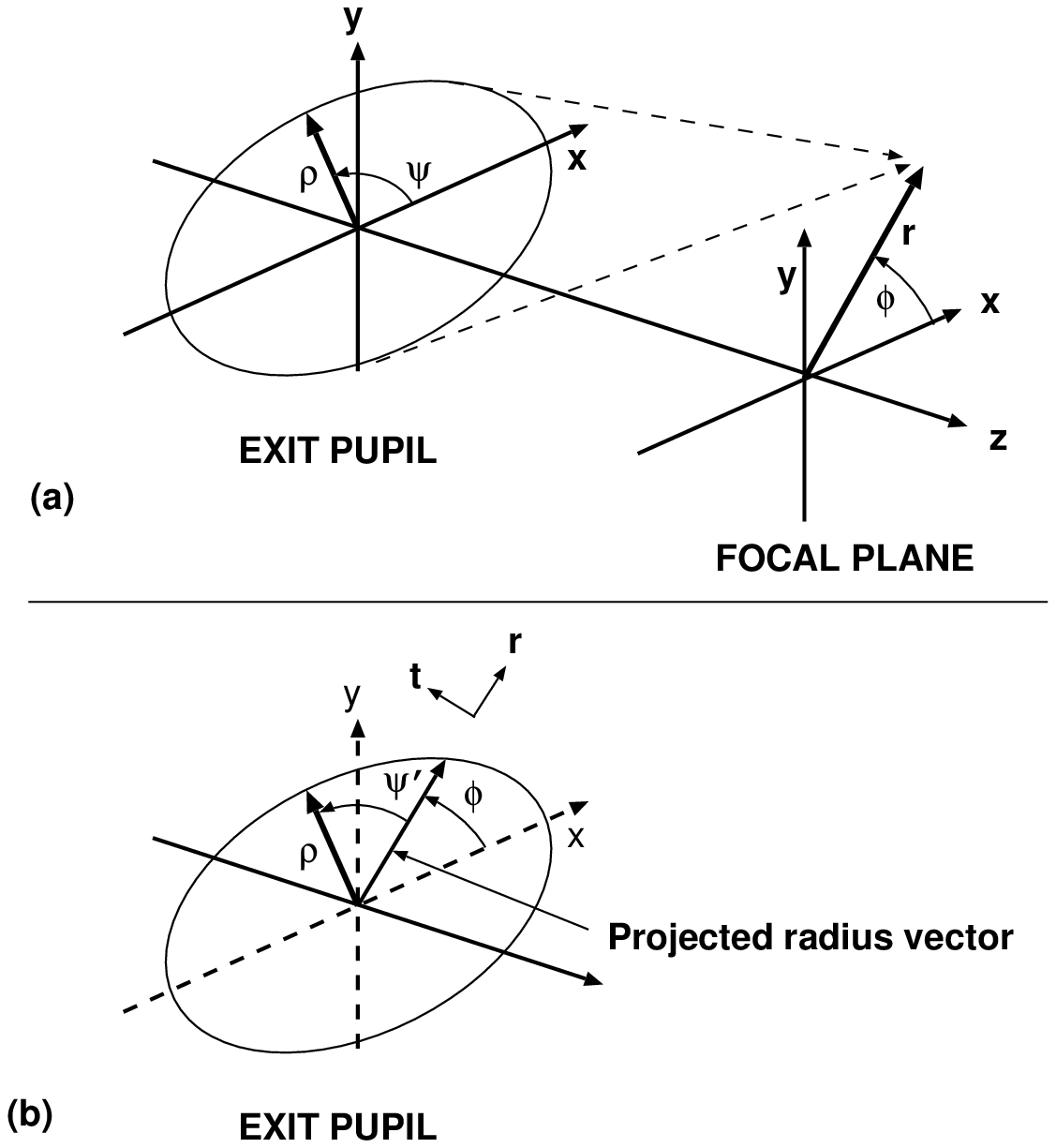,width=0.8\textwidth}
\caption{(a) Schematic diagram showing the orientation and coordinate system
definitions of the pupil plane ($\rho, \psi$) and focal plane ($r,\phi$).
Each point in the focal plane (e.g., at radius vector
$\vec{\textit{\textbf{r}}}$)
sees an image of the exit pupil.  A global Cartesian
$(x,y)$ system is illustrated.  (b) Another diagram of the
pupil plane, this time showing the projection of the radius vector
{\bf r} onto the pupil plane.
One can define local pupil axes (r,t) that
are aligned parallel and perpendicular to the projected radius vector
and that are rotated by $\phi$ relative to the global $(x,y)$ axes.
Pupil coordinates in this system are given by $\rho$ and
$\psi' = \psi - \phi$.}
\label{fig:geom}
\end{figure}

Figure \ref{fig:geom} shows the geometry of the exit pupil and focal plane.
The wavefront errors that give rise to optical aberrations can be
characterized in terms of Zernike polynomials,
which decompose the wavefront shape into 
orthogonal functions of the exit pupil polar coordinates
($\rho$,$\psi$).  At a particular point in the focal plane, one can write:
\begin{equation}
W(\rho,\psi) = \sum_l\sum_s
[A_{ls}\cos s\psi +
B_{ls}\sin s\psi]R_l^s(\rho),
\label{eq:wavefront}
\end{equation}
where $W$ is the 
wavefront error (in units, say, of microns or waves),
$R_l^s(\rho)$ is a Zernike radial polynomial, and
$A_{ls}$,$B_{ls}$
are Zernike aberration coefficients \citep[see, e.g.,][]{born00}.
Indices $l$ and $s$ are restricted to combinations where $l+s$ is even
and $s \le l$.
(For reasons that will
be apparent below, $l$ and $s$ are used here for
pupil polynomial indices, while
the more conventional $n$ and $m$ indices are reserved
for the focal plane.)  

The Zernike aberration coefficients $A,B$ are themselves a function 
of polar coordinates ($r,\phi$) in the focal plane, so in general
one can write them as $A_{ls}(r,\phi)$ and $B_{ls}(r,\phi)$.  The goal
of this paper is to develop efficient
polynomial expressions for these coefficients.

The orientation of the pupil coordinate system with respect to
the focal plane coordinates needs to be specified.
Normally, the pupil plane $x$ and $y$ axes are aligned with those of the
focal plane, forming a
global Cartesian system.  However, for an axisymmetric system,
a more natural pupil coordinate system is one in which
the angle $\psi$ is replaced with $\psi'=\psi-\phi$ at each location
in the focal plane such that the axes of
the pupil system are always oriented in the radial and tangential directions
relative to the optical axis \citep[e.g.,][]{braat13}.  In this case, the
aberration coefficients depend only on radial distance from the symmetry axis, 
and the patterns can be characterized with 1$-$d radial polynomials.
If the system is
perturbed from a state of axisymmetry, e.g., as happens when the
optical elements are misaligned or the corrector contains a 
non-axisymmetric element such as an atmospheric dispersion compensator 
(ADC), this axisymmetric formulation is no longer applicable.
However, the system still retains a high degree of symmetry, and thus
it seems desirable to seek a more general formulation for
the aberration pattern that still retains the radial/tangential
property such that it
reduces to the axisymmetric form in the limit of zero perturbation.

\begin{figure}
\centering
\epsfig{file=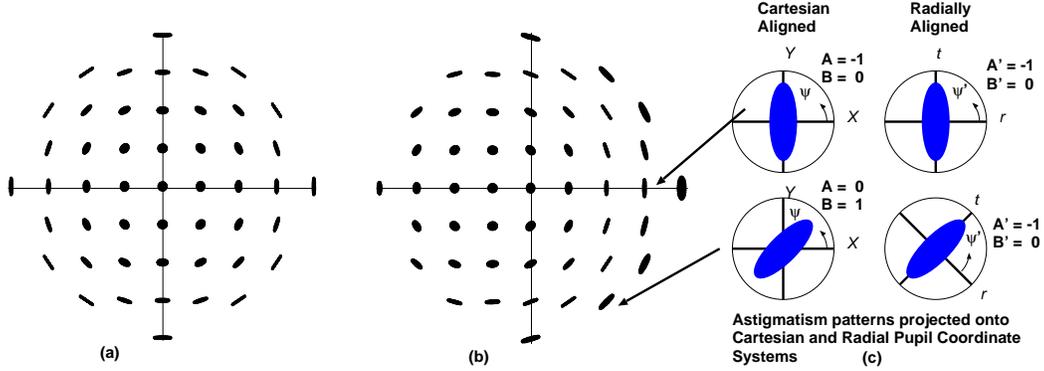,width=0.9\textwidth}
\caption{Astigmatism patterns as a function of
location in the focal plane for (a) aligned 2-mirror telescope and
(b) misaligned telescope with secondary mirror displaced and tilted.
Right panel (c) shows the projection of astigmatism pattern onto Cartesian
($A, B$) and radially aligned ($A', B'$) pupil axes.
The senses of the pupil azimuth angles $\psi$ and $\psi'$ are also shown.
The units are arbitrary.}
\label{fig:misalign}
\end{figure}

Figure \ref{fig:misalign} illustrates the situation.
This figure shows the astigmatism pattern at the focal plane of
a typical two-mirror telescope.  (The telescope is defocused slightly to
highlight the orientation of the pattern.)
In Fig.\ \ref{fig:misalign}(a), the telescope is properly
aligned, and the pattern is axisymmetric.
In Fig.\ \ref{fig:misalign}(b), the secondary mirror is offset, then tilted to
remove first-order coma; the astigmatism pattern is now perturbed, but
it still retains a high degree of symmetry.  In Fig.\ \ref{fig:misalign}(c),
the values for the aberration coefficients at two locations in the
focal plane are shown schematically when expressed
using both a Cartesian-aligned pupil coordinate system (left; $A$ and $B$)
and a radially-aligned system (right; $A'$ and $B'$).
In the former, the astigmatism amplitude oscillates between the
$A$ and $B$ terms
as one moves about the focal plane, even if the total amplitude
is constant.  In the latter, the amplitude is almost entirely in the $A'$
(radial) component, thus providing a clearer measure of the perturbation
pattern.

Multiple formulations for the dependence of aberration coefficients
on focal plane
coordinates have appeared in the literature
over the years, primarily in connection
with the problem of aligning telescopes with wide-field focal planes,
\citep[e.g.,][]{shack80,mcleod96,schech11}.
One approach is to express the aberration coefficients themselves
as a function of 
focal plane coordinates using the same Zernike polynomial basis
\citep[e.g.,][]{kwee93}.  Although such an expansion is straightforward to 
construct, the aberration coefficients are still
referenced to pupil axes that are aligned with a global Cartesian
system across the focal plane.

One can easily apply a rotation to the equations for the aberration
coefficients when expressed in a Cartesian-aligned pupil system in order
to convert them to a radially-aligned system, but 
the resulting expansions are no longer of Zernike form.  
Various alternative expansions using Zernike polynomials or products
of Zernike polynomials
have also been proposed \citep[e.g.,][]{agurok98,manuel09} but these
are still not quite of
the desired form.  \citet{gray15} came the closest, developing
expansions in the desired form up to eighth order in radius and third
order in azimuth by using methods 
of Geometric Algebra, but the resulting expressions are not easily
generalized to arbitrary order.

This difficulty suggests approaching the problem from a different direction.
From Eq. (\ref{eq:wavefront}),
one can show that, for a particular $l$, $s$ combination, the
quantity $A_{ls} + iB_{ls}$, expressed as a function of ($r,\phi$),
is a field of spin-weight $s$.
Such a field can can be derived from a scalar field by
successive application of spin raising operators (a
process analogous to the use of
angular momentum raising operators in quantum mechanics.)  This approach turns out to be fruitful.

Spin-weighted functions are not commonly used in optics, so a short
review is presented here.
\citep[See also] []{newman66,goldberg67,castillo92}.
Consider a local 
Cartesian system $(x,y)$ centered on some particular location in the
focal plane with basis vectors ${\bf e_1}, {\bf e_2}$ forming a 
right-handed system. A function ${}_sF(x,y)$ is said to have spin-weight
$s$ if, 
under a counterclockwise rotation of ${\bf e_1},\bf{e_2}$
by angle $\phi$ (i.e., 
${\bf e_1}' + i{\bf e_2}' = e^{-i\phi}[{\bf e_1} + i{\bf e_2}]$), the field 
in the rotated coordinate frame
transforms as ${}_sF' = e^{-is\phi}{~}_sF$.  (The sign convention is
chosen to match that of the CMB polarization literature.)
Thus, if one rewrites the coefficients of each term in Eq.~(\ref{eq:wavefront})
in the form
\begin{equation}
A\cos s\psi +B\sin s\psi = {1\over 2}(A-iB)e^{is\psi} + 
{1\over 2}(A+iB)e^{-is\psi},
\label{eq:complex}
\end{equation}
the coefficients of the terms on the right side have
spin-weights $-s$ and $+s$ respectively.

As an example, consider an optical system for which the
wavefront error includes terms corresponding
to astigmatism.
At a particular point in the focal plane $(x,y)$,
the two relevant terms for the wavefront error
are $A_{22}\rho^2\cos 2\psi$ and $B_{22}\rho^2\sin 2\psi$ 
\cite[Z6 and Z5 in the nomenclature of ][]{noll76}.
The complex quantity formed from the two coefficients,
${}_2F = A_{22}+iB_{22}$, expressed as a function of focal plane coordinates
$(x,y)$, has spin-weight $s=+2$.
To demonstrate this property, let $(x',y')$ be another pupil coordinate
system centered at this point and rotated counterclockwise
with respect to the $(x,y)$ system by an angle $\phi$.
In the new coordinate system, the complex coefficient should be given
by ${}_2F'=A_{22}'+iB_{22}' ={}_2Fe^{-2i\phi}$.
Equating the real and imaginary parts, the
coefficients transform as:
\begin{IEEEeqnarray}{rCl}
A_{22}'&=& A_{22}\cos 2\phi + B_{22}\sin 2\phi,
\IEEEyesnumber\IEEEyessubnumber*\\
B_{22}'&=& B_{22}\cos 2\phi - A_{22}\sin 2\phi
\end{IEEEeqnarray}
or
\begin{IEEEeqnarray}{rCl}
A_{22}&=& A'_{22}\cos 2\phi - B'_{22}\sin 2\phi,
\IEEEyesnumber\IEEEyessubnumber*\\
B_{22}&=& B'_{22}\cos 2\phi + A'_{22}\sin 2\phi.
\end{IEEEeqnarray}
The wavefront error itself at that point is given by:
\begin{IEEEeqnarray}{rCl}
W&=& A_{22}\rho^2\cos 2\psi + B_{22}\rho^2\sin 2\psi
\IEEEyesnumber\IEEEyessubnumber*\\
 &=& A'_{22}\rho^2\cos 2(\psi-\phi) + B'_{22}\rho^2\sin 2(\psi-\phi) \\
 &=& A'_{22}\rho^2\cos 2\psi' + B'_{22}\rho^2\sin 2\psi',
\end{IEEEeqnarray}
where $\psi' = \psi-\phi$.  Thus, the form of the wavefront aberration
is the same in the rotated frame as it is in the original frame.
If one rotates by $180^{\circ}$ degrees, the coefficients themselves
in the new frame are
unchanged from the old frame.  More generally, for any field of spin-weight
$s$, a local rotation of $360^\circ/s$ leaves the coefficients unchanged.

A key property of spin-weighted fields is that a field of positive spin-weight
$s$ can be generated from a scalar/pseudo-scalar
field ${}_0F = S-iK$ (with functions
$S$ and $K$ being real) by $s$ 
applications of the spin raising operator.  In Cartesian coordinates,
each step is of the form:
\begin{equation}
{}_{s+1}F = \edh{~}_sF=-\biggl[{\partial\over\partial x} 
+ i{\partial\over\partial y}\biggr]{~}_sF.\label{eq:edth}
\end{equation}
\citep[The symbol ``$\,\edh$\,'' was first used for this purpose
by][]{newman66}.  This result can be verified by examining the rotational
properties of $\,{}_{s+1}F$.
The real and imaginary parts correspond to 
curl-free and divergence-free fields respectively, and, by analogy to 
electromagnetism, this formulation
is sometimes referred to as E/B mode decomposition, a terminology that
will be used here as well.   For 
negative $s$, one takes the complex conjugate of Eq.~(\ref{eq:edth})
to create a spin  lowering function $\bar{\edh}$.  For simplicity,
only $s>=0$ will be considered in what follows, since the equations for
negative $s$ contain no additional information.

Such an approach is commonly used in problems involving
spin-weight 1 fields, such as atmospheric distortion \citep{bernstein17},
and spin-weight 2 fields, such as 
gravitational radiation \citep{newman66}, weak lensing
\citep{stebbins96}, and Cosmic Microwave Background
polarization \citep{kam97,zald97}.
For the spin-weight 2 problems, one normally uses spherical coordinates and 
expands the potentials using spherical harmonics.  Here, the natural 
coordinates are polar ($r,\phi$), and the preferred choice for
the radial functions is, again, Zernike radial polynomials
\citep[although other choices -- e.g. Bessel function -- are 
possible;][]{trevino13}.
One can transform from Cartesian to polar coordinates as follows.  Let:
\begin{IEEEeqnarray}{rCl}
{}_sF' &=& e^{-is\phi}{~}_sF,\IEEEyesnumber\IEEEyessubnumber*\\[1ex]
{}_{s+1}F' &=& e^{-i(s+1)\phi}{~}_{s+1}F, \\[1ex]
\biggl[{\partial\over\partial x} + i{\partial\over\partial y}\biggr]{~}_s F &=&
e^{i\phi}\biggl[{\partial\over\partial r} +
{i\over r}{\partial\over\partial\phi}\biggr] {~}_sF.
\end{IEEEeqnarray}
Upon substituting these expressions into Eq.~(\ref{eq:edth}), one
finds that the raising and lowering operators in polar coordinates
are given by \citep{castillo92}:
\begin{IEEEeqnarray}{rCl}
\edh {~}_sF &=& -\Biggl[{\partial\over\partial r} - {s\over r} + 
{i\over r}{\partial\over \partial\phi}\Biggr] {~}_sF,
\IEEEyesnumber\IEEEyessubnumber*\label{eq:rtraise} \\
\bar{\edh} {~}_sF 
&=& -\Biggl[{\partial\over\partial r} + {s\over r} - 
{i\over r}{\partial\over \partial\phi}\Biggr] {~}_sF.\label{eq:rtlower}
\end{IEEEeqnarray}

Common Zernike wavefront terms in Eq.~(\ref{eq:wavefront}) can be classified as
spin-weight  0 (defocus; spherical aberration), spin-weight 1
(distortion, coma), spin-weight 2 (astigmatism), spin-weight 3 (trefoil),
and spin-weight 4 (quadrafoil).  Generically, the coefficients for each
type of aberration are given by:
\begin{equation}
A'_{ls} + iB'_{ls} = {~}_sF' = \edh^s (S_{ls} - iK_{ls}),
\label{eq:generic}
\end{equation}
where the the notation $\edh^s$ means $s$ applications of the spin-raising
operator (going from spin-weight 0 to spin-weight $s$)
and the prime ($'$) again
indicates that the pupil coordinate system ($\rho,\psi'$,
where $\psi' = \psi-\phi$) is
rotated to align the axes in the radial and tangential directions at
any location in the focal plane.  Each aberration type
(e.g., third-order coma) corresponds to a particular pair of $(l,s)$
values (e.g., $l=3, s=1$) for which there is a corresponding pair
of scalar potentials $S_{ls}, K_{ls}$.  Given these potentials, the
aberration coefficients are derived as follows:

Spin-weight 0:
\begin{IEEEeqnarray}{rCl}
A'_{l0}(r,\phi) &=& S_{l0}(r,\phi),\IEEEyesnumber\IEEEyessubnumber* \\
B'_{l0}(r,\phi) &=& K_{l0}(r,\phi); \text{\ \ \ (not used)}
\end{IEEEeqnarray}

Spin-weight 1:
\begin{IEEEeqnarray}{rCl}
-A'_{l1}(r,\phi) &=& {\partial S_{l1}(r,\phi)\over\partial r} + {1\over r}
{\partial K_{l1}(r,\phi)\over\partial\phi},
\IEEEyesnumber\IEEEyessubnumber*\label{eq:spin1a} \\
-B'_{l1}(r,\phi) &=& {1\over r}{\partial S_{l1}(r,\phi)\over\partial\phi} -
{\partial K_{l1}(r,\phi)\over\partial r};\label{eq:spin1b}
\end{IEEEeqnarray}
etc. The equations for spin-weight
2 and greater are straightfoward to write down but
are unwieldly and will not be used directly in that form anyway.
It should be noted that the spin-weight 1 equations are a particular form of 
a Helmholtz decomposition.

The reason for starting with scalar fields $S$ and $K$ is that, being
spin-weight 0, each
can be expanded independently as a function of focal plane coordinates
using ordinary Zernike polynomials.  It is simplest to work with the Zernike
polynomials expressed using complex
notation.  Let
$C_n^m(r,\phi) = R_n^m(r)e^{im\phi},$ where any normalization,
such as the Noll convention, is omitted; the convention here for negative
angular index is that $R_n^{-m} = R_n^m$. The radius is in units of $r_{max}$,
the radius of the focal plane field of view.  One has:
\begin{IEEEeqnarray}{rCl}
S_{ls} &=& \sum_n \sum_m
a^{~ls}_{nm}C_n^m(r,\phi),
\IEEEyesnumber\IEEEyessubnumber*\label{eq:scalara} \\
K_{ls} &=& \sum_n \sum_m
b^{~ls}_{nm}C_n^m(r,\phi),\label{eq:scalarb}
\end{IEEEeqnarray}
where the sums extends over $m=-n$ to $n$ but restricted to $m+n$ even;
$a^{~ls}_{nm}$ and $b^{~ls}_{nm}$ are complex coefficients with the
constraint that the coefficients are Hermitian on index $m$ -- i.e.,
coefficients of index $-m$ are the complex conjugates of
the coefficients of index $+m$.

Upon substitution of
Eqs.~(\ref{eq:scalara}) and (\ref{eq:scalarb}) into Eq.~(\ref{eq:generic}),
one ends up with terms of the form $\edh^s C_n^m(r,\phi)$.
One can define corresponding spin-weighted Zernike polynomials ${}_sC_n^m$
as follows:
\begin{equation}
{}_sC_n^m(r,\phi) = (-1)^s~\edh^s C_n^m(r,\phi) = {~}_sR_n^m(r) e^{im\phi}.
\label{eq:spindef}
\end{equation}
The functions $_sR_n^m(r)$ are spin-weighted Zernike radial polynomials.
A listing of the functions ${}_sC_n^m$ up to $n=5$ is given in Appendix
\ref{appendix:a}.
Note that the radial polynomials are of order $n-s$ in radius, so it
is sometimes convenient to label them with $j=n-s$.

Unlike spin-weighted spherical harmonics, the spin-weighted Zernike
polynomials (other than for $s=0$), at least in this form,
are not orthogonal on index $n$.
However, one can orthogonalize them as follows.
By inspection, the spin-weighted
radial polynomials can be combined to create regular Zernike radial
polynomials; in particular, one can write:
\begin{equation}
R_{n-s}^{m+s}(r) = \sum_{k<=n} {}_sc_{nk} {~}_sR_k^m(r),\label{eq:ck}
\end{equation}
where the ${}_sc_{nk}$ are numerical coefficients and
$k+n$ is constrained to be even.  For example, for $s=1$, this equation
becomes $R_{n-1}^{m+1} = ({}_1R_n^m - {}_1R_{n-2}^m)/(2n)$.
Thus, one can convert
the equations for the aberration coefficients from a sum over ${}_sR_n^m(r)$
to a sum (with different coefficients) over $R_{n-s}^{m+s}$, as follows:
\begin{equation}
(-1)^s(A'_{ls}+iB'_{ls}) = 
\sum_n\sum_m (a^{'~ls}_{nm} - i\,b^{'~ls}_{nm})
R_{n-s}^{m+s}(r)e^{im\phi},\label{eq:ortho1}
\end{equation}
where the coefficients $a^{'~ls}_{nm}$ and $b^{'~ls}_{nm}$ are again complex.
$R_{n-s}^{m+s}(r)e^{im\phi}$
is an orthogonal form for the spin-weighted Zernikes.
The constraints are now $-n \le m \le n-2s$,
$n-s >= 0$, $n+m$ even, and the coefficients are again Hermitian on index $m$.
This last constraint follows as a consequence of the fact that the
coefficients in Eq.~(\ref{eq:ck}) are independent of $m$.
This formulation has the additional advantage that it is easier to
implement than the original (nonorthogonal) equations.
The one drawback (should
it matter) of the orthogonal form is that the connection
of the $a^{'~ls}_{nm}$ and $b^{'~ls}_{nm}$ to the original
coefficients in the scalar equations (\ref{eq:scalara})
and (\ref{eq:scalarb}) is somewhat complicated.  Details are presented in
Appendix \ref{sec:orthogonalization}.

For computational purposes, the sums over complex functions can be
rewritten as sums over real functions.  Splitting the complex
coefficients into real and imaginary components (where $,c$ and $,s$ are
used to label
cosine and sine coefficients respectively):
\begin{IEEEeqnarray}{rLc}
a_{nm}^{'~ls} = a_{nm,c}^{'~ls} - ia_{nm,s}^{'~ls}\,,
\IEEEyesnumber\IEEEyessubnumber*\label{eq:anm} \\
b_{nm}^{'~ls} = b_{nm,c}^{'~ls} - ib_{nm,c}^{'~ls}\,,
\label{eq:bnm}
\end{IEEEeqnarray}
 one has:
\begin{IEEEeqnarray}{rLL}
(-1)^sA'_{ls} =&& \sum_n\sum_m [a^{'~ls}_{nm,c}\cos m\phi +
a^{'~ls}_{nm,s}\sin m\phi][R_{n-s}^{-m+s}(r)+R_{n-s}^{m+s}(r)][1-
\delta_{m0}/2]
\IEEEnonumber \\
&& +\sum_n\sum_m [-b^{'~ls}_{nm,c}\sin m\phi + b^{'~ls}_{nm,s}\cos m\phi]
[R_{n-s}^{-m+s}(r) - R_{n-s}^{m+s}(r)],
\IEEEyesnumber\IEEEyessubnumber*\label{eq:A} \\
(-1)^sB'_{ls} =&& \sum_n\sum_m [-a^{'~ls}_{nm,c}\sin m\phi + 
a^{'~ls}_{nm,s} \cos m\phi]
[R_{n-s}^{-m+s}(r)-R_{n-s}^{m+s}(r)]\IEEEnonumber\IEEEnosubnumber \\
&& - \sum_n\sum_m[b^{'~ls}_{nm,c}\cos m\phi + b^{'~ls}_{nm,s}\sin m\phi]
[R_{n-s}^{-m+s}(r)+R_{n-s}^{m+s}(r)][1-\delta_{m0}/2].
\IEEEnonumber\IEEEnosubnumber \\
&&\IEEEyessubnumber
\label{eq:B}
\end{IEEEeqnarray}
These equations are the desired result.  $\delta_{m0}$ is the Kronecker
delta.
The sums are now over $m=0$ to $n$, again restricted to 
$m+n$ even, $m \le n$, and $n-s \ge 0$.
The individual terms on the right side
are the even and (negative) odd components of the spin-weighted radial
polynomials with respect to $m$.
For $n-s < m+s$, the term $R_{n-s}^{m+s}(r)$ in Eqs.~(\ref{eq:A}) and
(\ref{eq:B}) is 0,
and the equations become two-fold degenerate between the
$a$ and $b$ coefficients;
for these terms, the $b$ coefficients (corresponding to the B mode)
are set to 0.  

To recap, these equations provide a mechanism to parametrize the aberration
coefficients ($A',B'$) (expressed in a radially aligned pupil coordinate
system) as a function of position in the focal plane.
Each aberration type is specified by the values of
$(l,s)$ (e.g., coma corresponds to $l=3,s=1$ while astigmatism corresponds
to $l=2,s=2$).  The number of terms on the right side of the equations
[each term specified by $(n,m)$] required to characterize a pattern across the
focal plane depends on the details of the optical system.  Each term
is a polynomial in radius of degree $n-s$ and harmonic in azimuth angle
of order $m$.  For each $(n,m)$ combination, there are 4 coefficients,
two for the E mode ($a$) and two for the B mode ($b$).
In practice, one might have a map (either from raytracing or from direct
measurement) of the ($A',B'$) aberration
coefficients for each aberration type as a function of
position in the focal plane, and the coefficients on the right side
of Eqs.\ (\ref{eq:A}) and (\ref{eq:B}) are determined either from a
least-squares fit to that map or, if coverage is uniform, from taking
the inner product of map with each term (since the terms are orthogonal).

Sometimes it is useful to know the average power for a particular
($A',B'$) combination over the focal plane.  This quantity can be computed
from Eqs.\ (\ref{eq:A}) and (\ref{eq:B}), along with the
orthogonality relations for Zernike polynomials, yielding:
\begin{equation}
\begin{split}
\langle A_{ls}^{'2}+B_{ls}^{'2}\rangle =& \sum_n\sum_m
\Bigl[(a_{nm,c}^{'ls})^2 + (a_{nm,s}^{'ls})^2 + 
(b_{nm,c}^{'ls})^2 + (b_{nm,s}^{'ls})^2\Bigr]\cr
& \times [1-\delta_{m0}/2]/[n-s+1]
\end{split}
\label{eq:power}
\end{equation}
Finally
for completeness, the average power in wavefront errors at a particular
place in the focal plane can be computed from Eq.\ (\ref{eq:wavefront}):
\begin{equation}
\langle W^2\rangle = \sum_l\sum_s{1\over 2}(A_{ls}^{'2}+B_{ls}^{'2})
(1+\delta_{s0})/(2l+2)
\label{eq:wavepower}
\end{equation}

It is worth noting that, for the case $s=1$, \cite{zhao07,zhao08}
derived essentially all the results presented here (albeit for a slightly
different application involving distortion and optics testing) but for
a coordinate system that is Cartesian-aligned.  Indeed, it is
possible to express the spin-weighted Zernikes in Eqs.\ (\ref{eq:A}) and
(\ref{eq:B})
using their ``S'' and ``T'' polynomials
combined with a rotation.
Appendix \ref{appendix:b} provides more detail.

\begin{figure}
\centering
\begin{tabular}{cc}
\epsfig{file=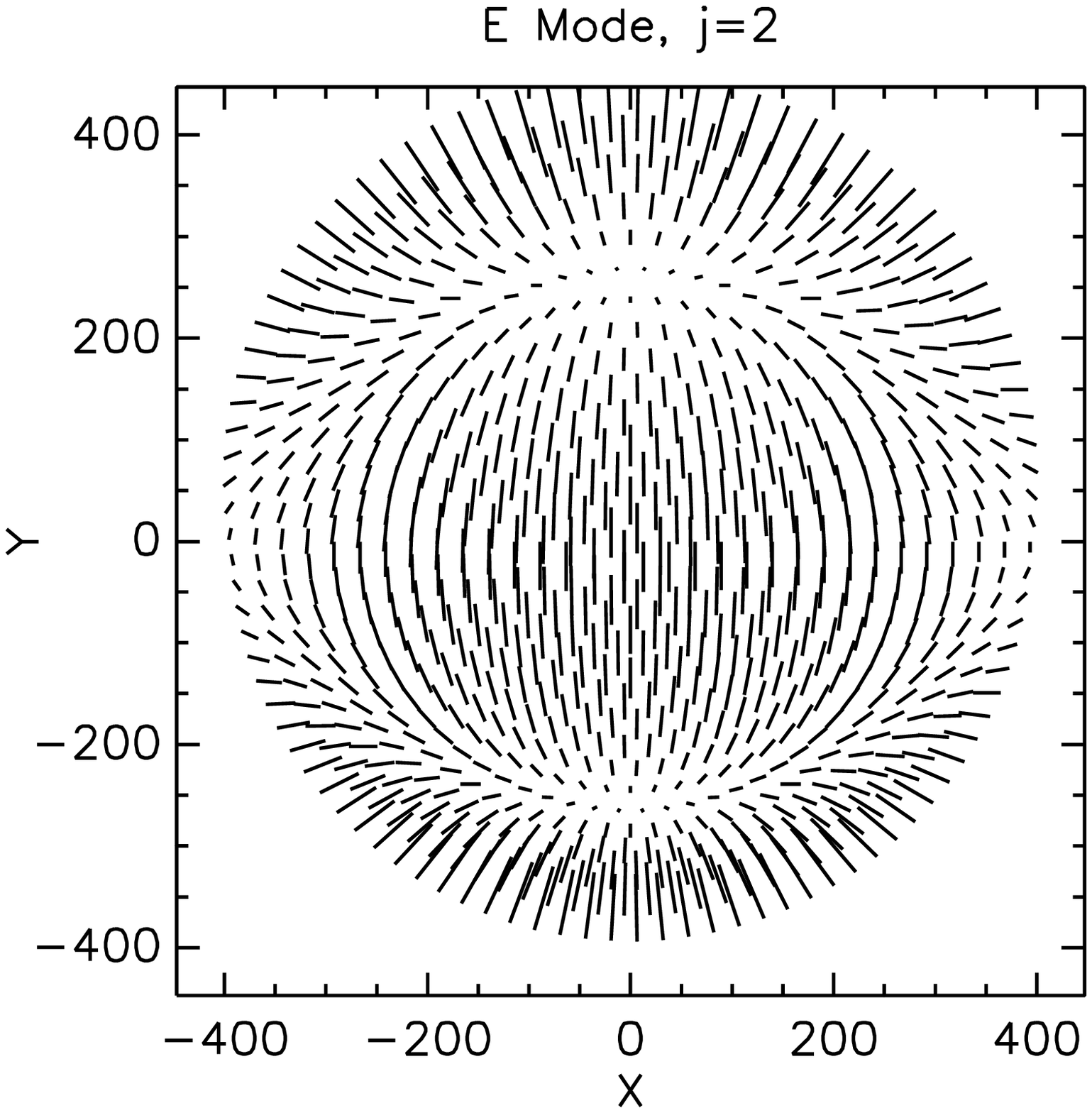,width=0.45\textwidth} &
\epsfig{file=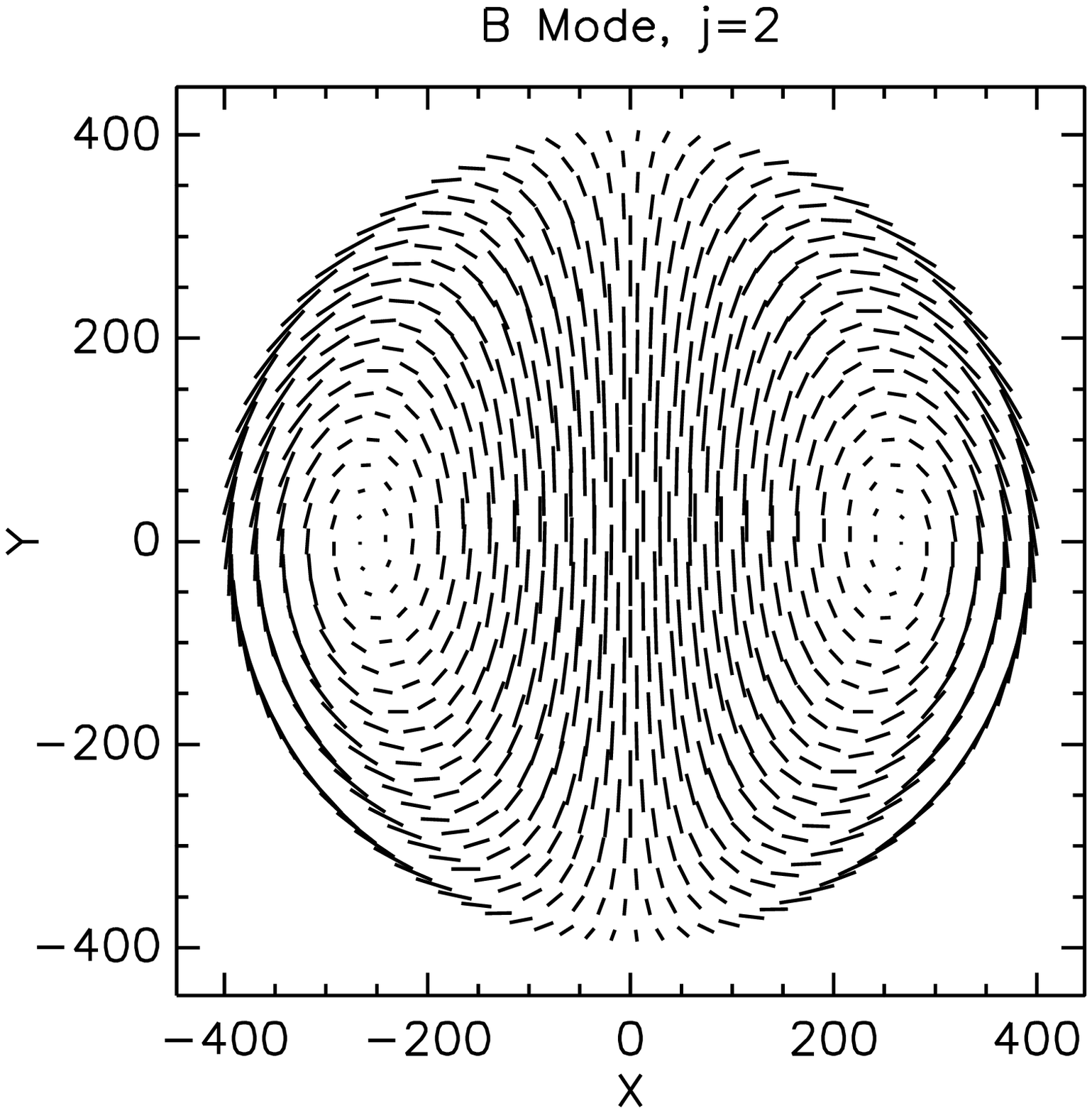,width=0.45\textwidth} \\
\end{tabular}
\caption{Aberration patterns for spin-weight 1 quantities (distortion,
coma) for $n=3$, ($j=2$), $m=1$.  Left: E-mode.  Right: B-mode.  The axes
are in units of mm (corresponding to the DESI focal plane described in
section \ref{sec:applications}.)  The lengths the tick marks are in
arbitrary units.}
\label{fig:modes}
\end{figure}

The difference between E-mode and B-mode patterns for one particular $(n,m)$
combination is shown
graphically  in Fig.\ \ref{fig:modes}.
[These modes correspond to $S_7$ and $T_8$ of \cite{zhao07,zhao08}, now
drawn side-by-side and at higher resolution.]  For the B-mode, two
centers of vorticity are clearly seen at $X=\pm 234$.  At the
periphery of the field, the E-mode vectors are primarily radial, while
the B-mode vectors are primarily tangential.  The equations for each
mode in this figure are given in Table \ref{tab:equations}.
\begin{deluxetable}{c c c c}
\tablecaption{Equations for Patterns in Fig.\ \ref{fig:modes}}
\tablehead{
\colhead{Mode} &
\colhead{Term} &
\colhead{$A'$} &
\colhead{$B'$} \\
&& \colhead{(Radial)} &
\colhead{(Tangential)}
}
\label{tab:equations}
\startdata
\noalign{\vskip 0.5em}
E & $a_{31,s}^{'~l1}$ & ($3r^2-1)\sin\phi$ & $(r^2-1)\cos\phi$ \\[1ex]
B & $b_{31,c}^{'~l1}$ & $-(r^2-1)\sin\phi$ & $-(3r^2-1)\cos\phi$ \\[1ex]
\enddata
\end{deluxetable}

Spin-weighted functions offer at least two benefits for parametrizing
aberration coefficients compared to previous approaches.
First, the pupil wavefront coefficients
$A'_{ls}$ and $B'_{ls}$ are referenced to radial and tangential
directions respectively
in the focal plane, and thus, for an axisymmetric system, all
of the coefficients that characterize the spatial dependence are zero
except for the $a^{'~ls}_{mn,c}$ terms with $m=0$.
As is shown in Appendix \ref{appendix:b}, even for nonaxisymmetric systems,
the use of radial and tangential
pupil coordinates results in equations that are more symmetrical than
those using Cartesian pupil coordinates.
(The precise choice for the center of the polar
coordinate system is not important - picking a different center would
result in a modification of all coefficients, but 
the form for all the equations would remain unchanged.)
Second, the $b^{'~ls}_{nm}$ coefficients, corresponding to the
B mode ($K$) in Eq. (\ref{eq:generic}), are usually negligible
unless the misalignment patterns are induced by multiple, sufficiently
strong, perturbations.  Thus, spin-weighted functions offer an efficient
way to represent aberration patterns.

\clearpage
\section{Applications} \label{sec:applications}

Two applications will be considered here -- one involving the misalignment
of a telescope, and the second being the mapping of
distortion patterns.

\subsection{Telescope Misalignment}

The first example is quite simple --
a Ritchey-Chr\'etien (RC) telescope in which the secondary
mirror has been displaced and tilted.  (Similar examples are commonly
analyzed in the literature.)  Telescope parameters are given
in Table \ref{tab:telescope}.  \cite[The unperturbed parameters
correspond to an example used by ][]{schroeder87}.  An RC design is
optimized to have no third-order coma when properly aligned;
the next highest aberration, astigmatism, is intrinsic to the design and has
a pattern like that shown in Fig.\ \ref{fig:misalign}, left side.
This aberration corresponds to the term $j=2, m=0$.

The telescope is misaligned by offseting the secondary mirror
from the optical axis of the primary mirror
by 5 mm and tilting the secondary to reobtain the best possible
images.  A conceptual repointing of the telescope is also done
to recenter the field.
Table \ref{tab:rc} list the major
residual perturbations induced by this misalignment, one
term for coma and two terms for astigmatism.
As is well known, a tilt and offset can be combined to eliminate coma at the
field center.  Here, the optimizer
retains a small amount of coma of constant amplitude and orientation
in order to balance astigmatism
and produce the best images.  These patterns are all E-mode.  

\begin{deluxetable}{l l r c}[b]
\tablecaption{Parameters of Ritchey Chretien Telescope\label{tab:telescope}}
\tablehead{
\colhead{Surface} &
\colhead{Parameter} &
\colhead{Value} &
\colhead{units}
}
\startdata
Primary & Diameter & $4000$ & mm \\
        & Radius of curvature & $-20000$ & mm \\
        & Conic constant & $-1.0417$ \\
        & Distance to secondary & $-7500$ & mm \\
Secondary & Radius of curvature & $-6666.67$ & mm \\
          & Conic constant & $-3.1728$ \\
          & Offset & $5$ & mm \\
          & Tilt & $0.119$ & deg \\
          & Distance to Focus & $10000$ & mm \\
Focal plane & Radius of curvature & $-2655$ & mm \\
            & Diameter & $240$ & mm \\
\enddata
\end{deluxetable}

\begin{deluxetable}{c c c r c c}[ht]
\tablecaption{Aberration Coefficients for Perturbed RC Telescope\label{tab:rc}}
\tablehead{
\colhead{Mode} &
\colhead{$j$} &
\colhead{$m$} &
\colhead{Amplitude} &
\colhead{$A'$} &
\colhead{$B'$} \\
&&&&
\colhead{(Radial)} &
\colhead{(Tangential)}
}
\startdata
\noalign{\vskip 0.5em}
\multicolumn{6}{l}{Coma: $l=3$, $s=1$} \\[0.5ex]
	E & 0& 1&   0.007 & $\cos\phi$ & $-\sin\phi$ \\[0.5ex]
\hline
\multicolumn{6}{l}{Astigmatism: $l=2$, $s=2$} \\[0.5ex]
	E & 0& 2&   -0.181 & $\cos 2\phi$ & $-\sin 2\phi$ \\[0.5ex]
	E & 1& 1&   0.962 & $r\cos\phi$ & $-r\sin\phi$ \\[0.5ex]
\enddata
\end{deluxetable}

\begin{figure}
\centering
\begin{tabular}{ccc}
\epsfig{file=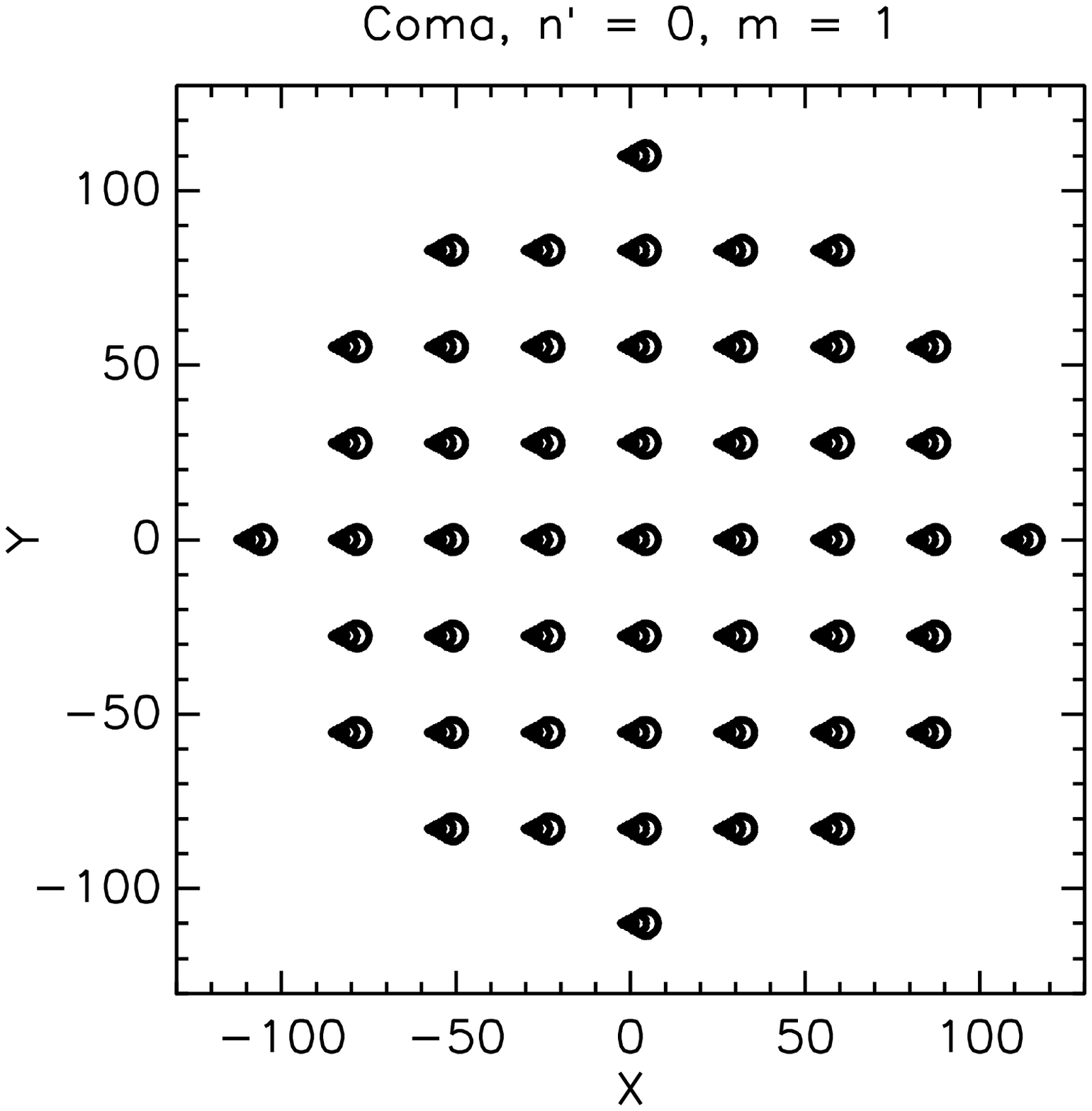,width=0.3\textwidth} &
\epsfig{file=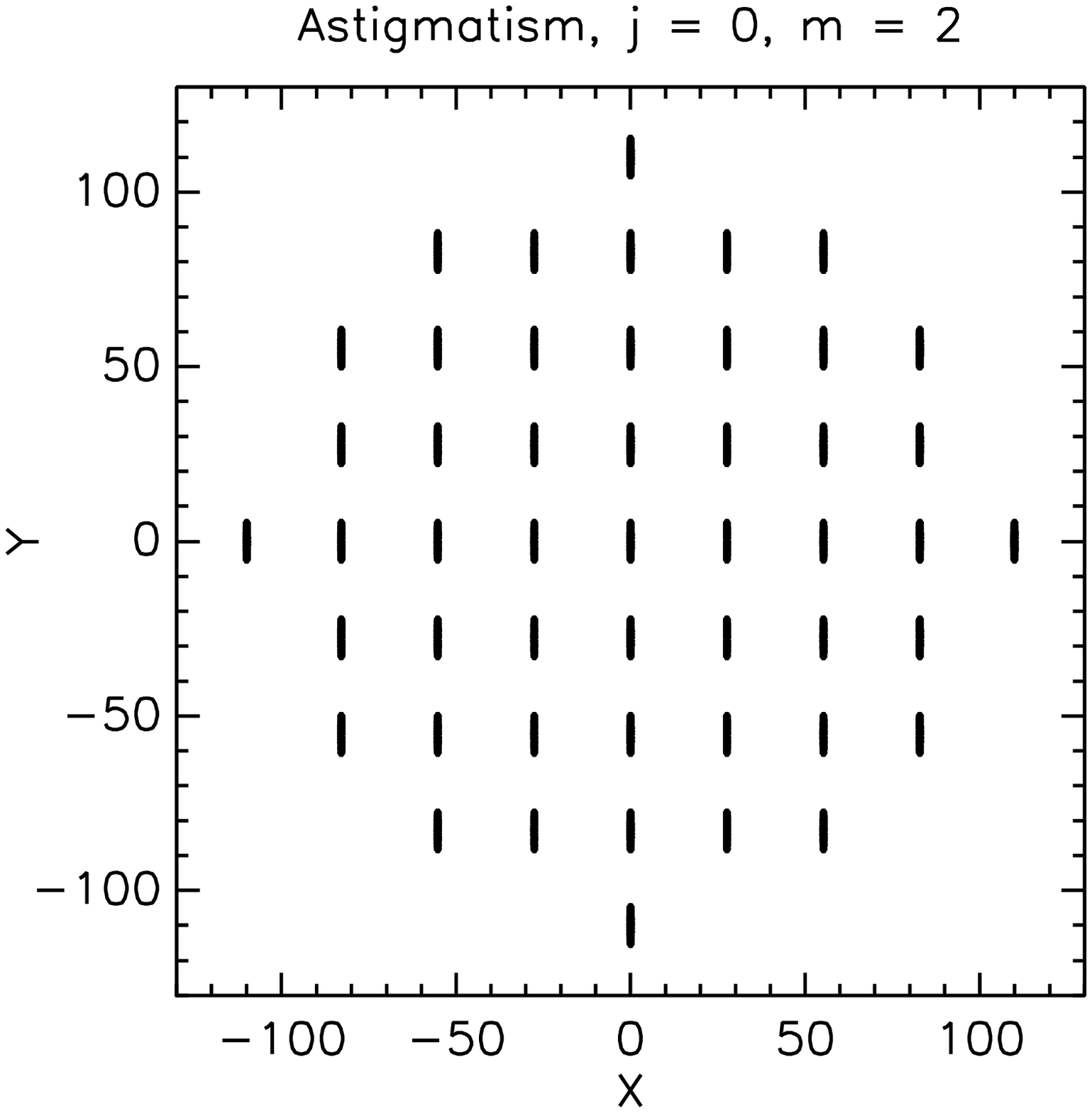,width=0.3\textwidth} &
\epsfig{file=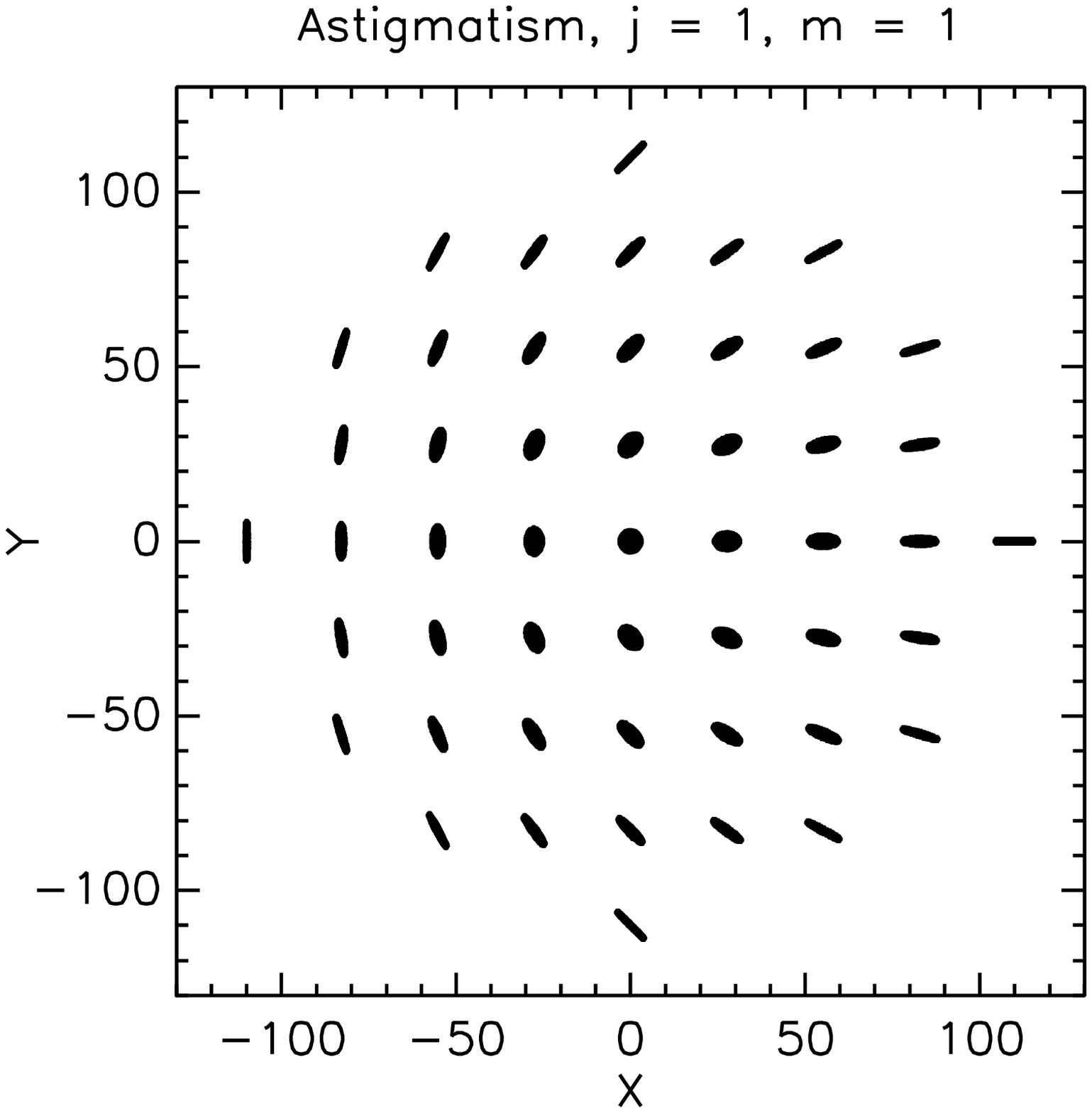,width=0.3\textwidth} \\
\end{tabular}
\caption{Focal plane aberration patterns for RC telescope with
offset, tilted secondary.  $X$ and $Y$ axis units are mm.  The aberration
patterns are not to scale and are for illustrative purposes only.}
\label{fig:rc}
\end{figure}

Figure \ref{fig:rc}
is a pictorial demonstration of the incremental patterns produced by each term.
(Once again, for astigmatism, the focus is shifted slightly to highlight
the orientation of the pattern.)

\subsection{DESI Distortion}

\begin{figure}[hb]
\begin{center}
\epsfig{file=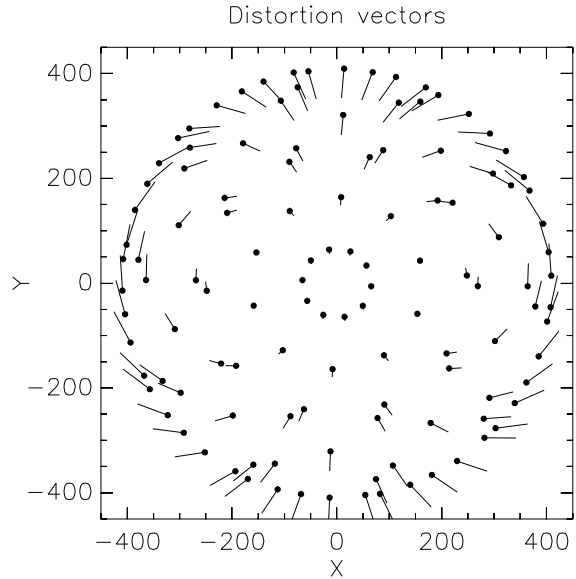,height=3in}
\caption{Incremental distortion in DESI corrector induced by
counter-rotating the lenses relative to the neutral position.
The $X$ and $Y$ axes are in units of mm.  The distortion ``whiskers''
have been scaled up in length by a factor $10^4$.}
\label{fig:distortion}
\end{center}
\end{figure}

The Dark Energy Spectroscopic Instrument (DESI),
currently under construction,
will have 5000 fiber positioners and will be used for a survey
measuring redshifts of 30 million galaxies \citep{desi16a,desi16b}.
It will be mounted at the prime focus of the
Mayall telescope, at Kitt Peak in Arizona.  A new, wide-field corrector
will provide a 3 degree diameter field of view.  In addition to
the fiber positioners, the DESI focal plane will be equipped with a
set of fixed fibers with good metrology that
will act as fiducial points to calibrate the geometry of the focal
plane.  Both
the positioner and fiducial fibers will be back-illuminated and imaged with a
Fiber View Camera (FVC), which will be
mounted near the primary mirror vertex.  Because
the FVC will view the focal plane through the corrector, it
will be necessary
to correct the positions measured by the FVC for distortion in the corrector.
The DESI corrector will incorporate an atmospheric
dispersion compensator (ADC), which will have two surfaces wedged relative
to the optical axis and which will introduce nonradial distortions.
Figure \ref{fig:distortion} shows the incremental
distortion induced by the ADC with the lenses counter-rotated relative
to the neutral position.  It is seen that
``vorticity'' is induced by the ADC.  Thus, both $E$ and $B$ modes
are needed to fully describe the pattern.  The $B$ mode, in fact,
is the same as that shown in Fig.\ref{fig:modes}.

A ``linear'' mapping of the sky (or object plane) to the focal (image)
plane would be as follows:
\begin{equation}
r = \theta/f,
\label{eq:scale}
\end{equation}
where $r$ is polar radius in the focal plane,
$\theta$ is the radial angle on the sky away from the field center, and $f$ is
the scale factor (arcsec mm$^{-1}$ or equivalent).  The azimuthal
angle $\phi$ is the same in both object and image planes.
In the presence of
distortion, the convention used here is for
$f$ to be chosen such that Eq.\ (\ref{eq:scale}) is exact at the center
and at edge of the focal plane.  (Any azimuthal dependence of the
scale factor is assumed to be averaged over.)
Thus, the radial distortion is zero at both the center and the edge
of the focal plane and non-zero in between, reaching over 6 mm
at the maximum.

The interpretation of the various terms, along with their amplitudes
at the edge of the field (in units of mm), is shown in Table \ref{tab:desi}.

\begin{deluxetable}{l c c r r c}
\tablecaption{Interpretation of full DESI parameters\label{tab:desi}}
\tablehead{
\colhead{Mode} &
\colhead{$j$} &
\colhead{$m$} &
\colhead{$\cos$} &
\colhead{$\sin$} &
\colhead{Description}
}
\startdata
E & 0 & 1 & 0.002 & 0.000 & Focal Plane decenter \\
           & 1 & 0 & -5.790 & -- & Scale factor \\
           & 2 & 1 & 0.000 & 0.092 & Optical axis decenter \\
           & 3 & 0 & 5.738 & -- & 3rd order radial distortion \\
           & 4 & 1 & 0.000 & 0.004 & Higher order decenter \\
           & 5 & 0 & 0.054 & -- & 5th order radial distortion \\
\hline
B & 0 & 1 & -- & --  & Not used \\
           & 1 & 0 & 0.000 & --  & Focal plane rotation \\
           & 2 & 1 & 0.111 & 0.000 & Vorticity \\
           & 3 & 0 & 0.000 & --  & Cubic spiral \\
           & 4 & 1 & 0.000 & 0.000  & Higher order vorticity \\
           & 5 & 0 & 0.000 & --  & Fifth order spiral \\
\enddata
\end{deluxetable}

Distortion corresponds to the $l=1$, $s=1$ terms in Eq.~(\ref{eq:wavefront}),
with a proportionality constant that is not of concern here.
Specifically, at a point ($r,\phi$),
distortion in the radial and tangential directions is given by
$\Delta r \propto A'_{11}$ and $\Delta t = r\Delta\phi \propto B'_{11}$
respectively.
(Note: for practical convenience, a local Cartesian system is defined
around the nominal
object center such that one axis is aligned with the radius vector;
$\Delta r$ and $\Delta t$ are actually measured as Cartesian
offsets in that system.)
The convention here is that these increments are the differences between
the actual location in the focal plane for the centroid of an object
and the location predicted
by the linear mapping as defined above.

Once again,
it is convenient to replace the radial index $n$ with $j = n-s$ and $s=1$,
since the radial polynomials are of this order.
In practice, only the $m=0$ and $m=1$ azimuthal terms are important for the
DESI distortion pattern, and the $b$ terms for $j=0$ are degenerate with
the $a$ terms and are set to 0.
The pattern predicted by raytracing is fully described by terms
up to $j=5$, with an rms error of 1 micron.  The B mode
contributes about 111 microns (approximately one fiber diameter)
at maximum rotation of the ADC.
The total number of coefficients included is 16 (although several
are essentially 0 and may be unnecessary once the actual corrector is
assembled and tested.)
For comparison, a more conventional Cartesian expansion requires
42 terms \citep[e.g.,][extended to 5th order]{anderson03}.
(For completeness, there are an additional two terms
used to define the center of the sky or FVC pattern; these are computed by
averaging the positions of all the fiducial spots in the FVC CCD,
with no further adjustment.)

\section{Conclusions}

A method has been presented in which Zernike aberration coefficients
are expressed in radially-oriented pupil coordinates so as to treat
axisymmetric and non-axisymmetric optical systems on an equal footing.
The method is general to arbitrary order in both azimuth and angular
order in the focal plane and to arbitrary order of aberration.

The algorithms for distortion are currently incorporated in the DESI
application ``PlateMaker'', which is used for measuring distortion
and mapping the focal plane to the sky.  They have already been applied to
data obtained with a prototype instrument ``protoDESI''
\citep{parker16,parker17}
to analyze the behavior and stability of the instrument.

Aberration patterns other than distortion
can be measured using wide-field exposures
of defocused ``donut'' images of stars, as is currently done for
the Dark Energy Camera \citep[DECam; ][]{flaugher15}
at CTIO \citep{rood14}.  Analysis of donut data from DECam using
the methods presented here is currently in
progress and will be reported elsewhere.  These methods will also be
of value  in the future for analyzing images from wide-field telescopes
such as LSST.

Finally, E/B mode decomposition may be of benefit for
analyzing aberration patterns of off-axis telescopes, where, for
certain designs, both E and B modes might be expected.

\acknowledgements

The author would like to thank Mike
Lampton and Albert Stebbins for many stimulating discussions that contributed
to the results presented here, and to Tod Lauer, Connie Rockosi, Paul
Martini, Liz Buckley-Geer, and the referee for carefully reading
the manuscript and providing
invaluable feedback that greatly improved its clarity.

This manuscript has been authored by
Fermi Research Alliance, LLC under Contract No. DE-AC02-07CH11359 with the
U.S. Department of Energy, Office of Science, Office of High Energy Physics.
The United States Government retains
and the publisher, by accepting the article for publication, acknowledges
that the United States Government retains a non-exclusive, paid-up,
irrevocable, world-wide license to publish or reproduce the published form
of this manuscript, or allow others to do so, for United States Government
purposes.

\clearpage
\appendix
\section{Spin-weighted Zernike Polynomials}
\label{appendix:a}
Table \ref{tab:zernike}
provides a list of the non-orthogonal form of
spin-weighted Zernike polynomials ${}_sC_n^m$
up to $n=s=5$.\footnote{By using Mathematica, the referee was able to
express these functions in closed form involving
the  use of Jacobi polynomials.}
Terms not listed are 0.  Note that all terms with $n-s<m+s$ are 0.

\begin{deluxetable}{L@{ = }L L@{ = }L L@{ = }L}[hb]
\tablecaption{Spin-Weighted Zernike Polynomials\label{tab:zernike}}
\tablehead{\multicolumn{6}{l}{\text{Spin-weight 0}}}
\startdata
C_0^0 & 1 & C_1^{-1} & re^{-i\phi} & C_1^1 & re^{i\phi}\\
C_2^{-2} & r^2e^{-2i\phi} & C_2^0 & 2r^2-1 & C_2^2 & r^2e^{2i\phi}\\
C_3^{-3} & r^3e^{-3i\phi} & C_3^{-1} & (3r^3-2r)e^{-i\phi} & 
C_3^1 & (3r^3-2r)e^{i\phi} \\
C_3^3 & r^3e^{3i\phi} & C_4^{-4} & r^4e^{-4i\phi} & 
C_4^{-2} & (4r^4-3r^2)e^{-2i\phi}\\
C_4^0 & (6r^4-6r^2+1) & C_4^2 & (4r^4-3r^2)e^{2i\phi} & 
C_4^4 &  r^4e^{4i\phi}\\
C_5^{-5} & r^5e^{-5i\phi} & C_5^{-3} & (5r^5-4r^3)e^{-3i\phi} &
C_5^{-1} & (10r^5-12r^3+3r)e^{-i\phi} \\
C_5^1 & (10r^5-12r^3+3r)e^{i\phi} & C_5^3 & (5r^5-4r^3)e^{3i\phi} &
C_5^5 & r^5e^{5i\phi}\\ [0.5ex]
\hline
\multicolumn{6}{l}{\text{Spin-weight 1}} \\
\hline
{}_1C_1^{-1} & 2e^{-i\phi} & {}_1C_2^{-2} & 4re^{-2i\phi} & {}_1C_2^0 & 4r\\
{}_1C_3^{-3} & 6r^2e^{-3i\phi} & {}_1C_3^{-1} & (12r^2-4)e^{-i\phi} & 
{}_1C_3^1 & 6r^2e^{i\phi}\\
{}_1C_4^{-4} & 8r^3e^{-4i\phi} & {}_1C_4^{-2} & (24r^3-12r)e^{-2i\phi} & 
{}_1C_4^0 & 24r^3-12r\\
{}_1C_4^2 & 8r^3e^{2i\phi} & {}_1C_5^{-5} & 10r^4e^{-5i\phi} & {}_1C_5^{-3} & 
(40r^4-24r^2)e^{-3i\phi}\\
{}_1C_5^{-1} & (60r^4-48r^2+6)e^{-i\phi} & {}_1C_5^1 & (40r^4-24r^2)e^{i\phi} &
{}_1C_5^3 & 10r^4e^{3i\phi}\\ [0.5ex]
\hline
\multicolumn{6}{l}{\text{Spin-weight 2}} \\
\hline
{}_2C_2^{-2} & 8e^{-2i\phi} & {}_2C_3^{-3} & 24re^{-3i\phi} &
{}_2C_3^{-1} & 24re^{-i\phi}\\
{}_2C_4^{-4} & 48r^2e^{-4i\phi} & {}_2C_4^{-2} & (96r^2-24)e^{-2i\phi} & 
{}_2C_4^0 & 48r^2\\
{}_2C_5^{-5} & 80r^3e^{-5i\phi} & {}_2C_5^{-3} & (240r^3-96r)e^{-3i\phi} &
{}_2C_5^{-1} & (240r^3-96r)e^{-i\phi}\\
{}_2C_5^1 & 80r^3e^{i\phi}\\ [0.5ex]
\hline
\multicolumn{6}{l}{\text{Spin-weight 3}} \\
\hline
{}_3C_3^{-3} & 48e^{-3i\phi} & {}_3C_4^{-4} & 192re^{-4i\phi} &
{}_3C_4^{-2} & 192re^{-2i\phi}\\
{}_3C_5^{-5} & 480r^2e^{-5i\phi} & {}_3C_5^{-3} & (960r^2-192)e^{-3i\phi} &
{}_3C_5^{-1} & 480r^2e^{-i\phi}\\ [0.5ex]
\hline
\multicolumn{6}{l}{\text{Spin-weight 4}} \\
\hline
{}_4C_4^{-4} & 384e^{-4i\phi} & {}_4C_5^{-5} & 1920re^{-5i\phi} &
{}_4C_5^{-3} & 1920re^{-3i\phi}\\
\hline
\multicolumn{6}{l}{\text{Spin-weight 5}} \\
\hline
{}_5C_5^{-5} & 3840e^{-5i\phi}\\
\enddata
\end{deluxetable}

\section{Orthogonalization of Spin-Weighted Functions}
\label{sec:orthogonalization}
\setcounter{equation}{0}
This section give the details of the orthogonalization process.
Specifically, the goal is to show that, if one
applies the spin-raising operator to an equation of the form given
by Eq.\ (\ref{eq:ortho1}), the resulting equation can be rewritten
to once again have the form of Eq.\ (\ref{eq:ortho1}) but with the spin-weight
$s$ incremented by 1.  By repeating this process incrementally after
each application of the spin-raising operator (starting with spin-weight 0),
one has effectively
achieved the orthgonalization process given by Eq.\ (\ref{eq:ck}).

As an initial matter,
Equation (\ref{eq:ortho1})
can be written in a slightly different form by shifting the angular
indices $m$ and $n$ as follows:
\begin{equation}
(-1)^s(A'_{ls}+iB'_{ls}) = 
e^{-is\phi}\sum_{n'}\sum_{m'} (a^{'~ls}_{n'm'} - i\,b^{'~ls}_{n'm'})
R_{n'}^{m'}(r)e^{im'\phi},\label{eq:ortho2}
\end{equation}
where $n'=n-s$ and $m'=m+s$.  In this form, the double sums are now symmetric
about $m'$ (i.e. $m'$ goes from $-n'$ to $n'$, $m'+n'$ even) and the sum
over $n'$ starts at 0.
The leading exponential is the term that rotates spin-weight $s$ quantities
from a Cartesian frame to a radial/tangential frame.  Thus, the double
sum gives the quantity $A+iB$, and the exponential converts
it to $A'+iB'$.  One twist, however,
is that the coefficients $a^{'}$ and $b^{'}$ are Hermitian on $m$, not $m'$.

To simplify notation, the right side of Eq.\ (\ref{eq:ortho2}) is rewritten
as follows, retaining just the $a^{'}$ coefficients and one value of $m'$:
\begin{equation}
\sum_{n'} a^{'}_{n'} R_{n'}^{m'}(r)e^{i(m'-s)\phi}.\label{eq:simplify}
\end{equation}
Applying the spin-raising operator, Eq.\ (\ref{eq:rtraise}),
gives the expression:
\begin{equation}
\sum_{n'} a^{'}_{n'}\biggl[{d\over dr}-{m'\over r}\biggr]R_{n'}^{m'}(r)
e^{i(m'-s)\phi}.\label{eq:raise}
\end{equation}
Notably, the radial part of this expression does not depend on the
spin-weight $s$.  The next step involves replacing the sum over the radial
functions (which now are not orthogonal) with a sum using different
coefficients over regular Zernike radial polynomials.
From \citet{lukosz63}, Eq.\ (A II.4b)
and \citet{jan14}, Eq.\ (8), one has:
\begin{equation}
\biggl[{d\over dr} - {m'\over r}\biggr]R_{n'}^{m'}(r) = 2n'R_{n'-1}^{m'+1} +
\biggl[{d\over dr} - {m'\over r}\biggr]R_{n'-2}^{m'}(r).\label{eq:deriv}
\end{equation}
Let:
\begin{equation}
{}_1R_{n'}^{m'} = {dR_{n'}^{m'}\over dr} - {m'\over r}R_{n'}^{m'},
\label{eq:spin1}
\end{equation}
where the notation is consistent with that used in Eq.\ (\ref{eq:spindef}).
By substituting Eq.\ (\ref{eq:spin1}) into Eq.\ (\ref{eq:deriv}), one gets:
\begin{equation}
{}_1R_{n'}^{m'} = 2n'R_{n'-1}^{m'+1} + {}_1R_{n'-2}^{m'}.
\end{equation}
One can rewrite this equation as:
\begin{equation}
R_{n'-1}^{m'+1} = ({}_1R_{n'}^{m'} - {}_1R_{n'-2}^{m'})/2n'.
\end{equation}
The difference equation can be solved easily to yield:
\begin{equation}
{}_1R_{n'}^{m'}=\sum_k (2k)R_{k-1}^{m'+1},\label{eq:sum}
\end{equation}
where the sum extends over $1 \le k \le n'$
subject to the constraint that $k$ has the same parity as $n'$.  By combining
Eqs.\ (\ref{eq:sum}) with (\ref{eq:spin1}) and substituting into
Eq.\ (\ref{eq:raise}), one now has a (double) sum over ordinary Zernike radial
polynomials.  The double sum can be simplified by introducing new
coefficients $a^{''}$ as follows:
\begin{IEEEeqnarray}{rLL}
a^{'}_{n'} &=& {a^{''}_{n'}\over 2n'} - {a^{''}_{n'+2}\over 2(n'+2)}
\IEEEyesnumber\IEEEyessubnumber*\label{eq:convert} \\[1ex]
a^{''}_{n'} &=& 2n'\sum_k a^{'}_k,
\label{eq:revert}
\end{IEEEeqnarray}
where the latter sum starts at $k=n'$ and extends to the upper limit of
the expansion over $a_{n'}$ with the constraint that $k$ has the same parity
as $n'$.
In one last step, the indices can be shifted yet again:
$n{''} = n'-1, m{''} = m'+1$,
and the spin-raised equivalent of Eq.\ (\ref{eq:simplify}) finally
has the form:
\begin{equation}
\sum_{n{''}} a^{''}_{n{''}}\, R_{n{''}}^{m{''}}(r)e^{i[m{''}-(s+1)]\phi}.
\label{eq:spinup}
\end{equation}
To summarize, the effect of applying the spin-raising operator is to
shuffle the coefficients (replacing a sum over $a^{'}$ with a sum over
$a^{''}$) and increase the spin-weight $s$ by 1.

The same equations also hold for the $b$ coefficients.  Because the
equations for these coefficients do not depend on $m$ (or $m'$),
the new coefficients $a^{''}$ and $b^{''}$ remain Hermitian on $m$.

\section{Comparison to other work}
\label{appendix:b}
\setcounter{equation}{0}
\cite{zhao07,zhao08} derived expansions for a distortion field
(i.e., spin-weight 1) that are essentially
identical to those here, except that they used a coordinate system
that is Cartesian-aligned.  The purpose here is to show how the two are
connected.  This connection will be demonstrated explicitly
for one term from the 2007 paper,
Table 4: $S_{16}$.  In the notation of that paper, this term is given by:
\begin{equation}
S_{16} = {1\over 2}[\hat{i}(\sqrt{2}Z_{11}+Z_{12})+\hat{j} Z_{13}].
\end{equation}
The Noll form of the Zernike polynomials is given by:
\begin{IEEEeqnarray}{rLL}
Z_{11} &=& \sqrt{5} {R\,}_4^0(r),
\IEEEyesnumber\IEEEyessubnumber* \\
Z_{12} &=& \sqrt{10} {R\,}_4^2(r)\cos 2\phi, \\
Z_{13} &=& \sqrt{10} {R\,}_4^2(r)\sin 2\phi,
\end{IEEEeqnarray}
with
\begin{IEEEeqnarray}{rLL}
{R\,}_4^0(r) &=& 6r^4-6r^2+1,
\IEEEyesnumber\IEEEyessubnumber* \label{eq:r40} \\
{R\,}_4^2(r) &=& 4r^4-3r^2.\label{eq:r42}
\end{IEEEeqnarray}

These equations represent the Cartesian $x$-axis and $y$-axis components of
a vector.  In the notation of the present paper, these components
are called $A$ and
$B$.  Thus, one has $S_{16} = \hat{i}A + \hat{j}B$, with:
\begin{IEEEeqnarray}{rLL}
A &=& {\sqrt{10}\over 2}[{R\,}_4^0(r) + {R\,}_4^2(r)\cos 2\phi],
\IEEEyesnumber\IEEEyessubnumber* \label{eq:16x} \\
B &=& {\sqrt{10}\over 2}{R\,}_4^2(r)\sin 2\phi. \label{eq:16y}
\end{IEEEeqnarray}
The conversion to radial and tangential components $A'$ and $B'$ is
given by:
\begin{IEEEeqnarray}{rLc}
A' &=&\phantom{Aa}A\cos\phi + B\sin\phi,
\IEEEyesnumber\IEEEyessubnumber* \\
B' &=&-A\sin\phi + B\cos\phi.
\end{IEEEeqnarray}
Substituting, one gets:
\begin{IEEEeqnarray}{rLc}
A' =& \sqrt{10}
[{R\,}_4^2(r) \sin 2\phi\cos\phi + {R\,}_4^0(r)\sin\phi -
{R\,}_4^2(r)\sin\phi\cos 2\phi],
\IEEEyesnumber\IEEEyessubnumber* \\
B' =& \sqrt{10}[-{R\,}_4^2(r)\sin 2\phi\sin\phi + {R\,}_4^0(r)\cos\phi -
{R\,}_4^2(r)\cos\phi\cos 2\phi].
\end{IEEEeqnarray}
Making use of relations for the products of trigonometric functions, one
finally gets:
\begin{IEEEeqnarray}{rLc}
A' =& \phantom{Aa}\sqrt{10}[{R\,}_4^0(r) + {R\,}_4^2(r)]\cos\phi,
\IEEEyesnumber\IEEEyessubnumber*\label{eq:c16a} \\
B' =& -\sqrt{10}[{R\,}_4^0(r) - {R\,}_4^2(r)]\sin\phi.
\label{eq:c16b}
\end{IEEEeqnarray}
By comparison with Eqs.~(\ref{eq:A}) and (\ref{eq:B}), one sees that $S_{16}$
corresponds to the $a_{nm,c}^{'~ls}$ term
with $s=1, n=5, m=1$.  (Likewise, $S_{17}$ corresponds to the
$a_{nm,s}^{'~ls}$ term, $T_{16}$ corresponds to the $b_{nm,c}^{'~ls}$ term,
and $T_{17}$ corresponds to the $b_{nm,s}^{'~ls}$ term.)
Note that the form of Eqs. (\ref{eq:c16a}) and
(\ref{eq:c16b}) is more symmetrical than that of (\ref{eq:16x}) and
(\ref{eq:16y}) in that it does not mix together terms of different harmonic
order $m$.


\clearpage

\begin{thebibliography}{dummy}

\bibitem[Agurok(1998)]{agurok98}
Agurok, I.\ 1998, Proc.\ SPIE, 3430, 80

\bibitem[Akiyama(2008)]{akiyama08}
Akiyama, M. et~al.\ 2008,
Proc. SPIE, 7018, 2

\bibitem[Anderson \& King(2003)]{anderson03}
Anderson, J., \& King, I. R.\ 2003,
PASP, 115, 113

\bibitem[Bernstein et~al.(2017)]{bernstein17}
Bernstein, G. et al.\ 2017, PASP, 129, 4503

\bibitem[Born \& Wolf(2000)]{born00}
Born, M., \& Wolf, E.\ (2000), ``Principles of Optics (7th Edition)"
(Cambridge University Press: Cambridge)

\bibitem[Braat \& Janssen(2013)]{braat13}
Braat, J. J.~M., \& Janssen, A. J. E.~M. \ 2013,
JOSA A, 30, 1213

\bibitem[DESI Collaboration(2016a)]{desi16a}
DESI Collaboration 2016, arXiv:1611.00036

\bibitem[DESI Collaboration(2016b)]{desi16b}
DESI Collaboration 2016, arXiv:1611.00037

\bibitem[Fagrelius et~al.(2016)]{parker16}
Fagrelius, P. et al, 2016,
Proc. SPIE, 9908,7

\bibitem[Fagrelius et~al.(2017)]{parker17}
Fagrelius, P. et~al, 2017,
arXiv:1710.08875

\bibitem[Flaugher et~al.(2015)]{flaugher15}
Flaugher, B. et al.\ 2015,
\aj, 150,150

\bibitem[Goldberg et~al.(1967)]{goldberg67}
Goldberg, J.~N. \ et al.\ 1967,
J. Math. Phys., 8,2155

\bibitem[Gray \& Rolland(2015)]{gray15}
Gray, R.~W., \& Rolland, J.~P. 2015,
JOSA A, 32, 1836

\bibitem[Hamana(2013)]{hamana13}
Hamana, T. et al.\ 2013,
PASJ, 65, 104

% \bibitem[Ivezic et~al.(2008)]{lsst08}
% Ivezic, Z. et al.\ 2008, arXiv:0805.2366

\bibitem[Janssen(2014)]{jan14}
Janssen, A. J. E. M.\ 2014, J. Opt. Soc. Am., 31, 1084

\bibitem[Jarvis et~al.(2008)]{jarvis08}
Jarvis, M., Schechter, P., \& Jain, B.\ 2008, arXiv:0810.0027

\bibitem[Jee \& Tyson(2011)]{jee11}
Jee, M.~J., \& Tyson, J.~A.\ 2011,
PASP, 123, 596

\bibitem[Kamionkowski et~al.(1997)]{kam97}
Kamionkowski, M., Koposov, A., \& Stebbins, A.\ 1997,
Phys. Rev. D, 55, 7368

\bibitem[Kent et~al.(2016)]{kent16}
Kent, S. M. et al.\ 2016,
Proc. SPIE, 9908, 8

\bibitem[Kwee \& Braat(1993)]{kwee93}
Kwee, I.~W., \& Braat, J. J.~M.\ 1993,
Pure Appl. Opt., 2, 21

\bibitem[Lukosz(1963)]{lukosz63}
Lukosz, W.\ 1963,
Opt. Acta, 10, 1

\bibitem[Manuel(2009)]{manuel09}
Manuel, A.~M. 2009, PhD thesis, U. Arizona

\bibitem[McLeod(1996)]{mcleod96}
McLeod, B.~A.\ 1996,
PASP, 2, 217

\bibitem[Newman \& Penrose(1966)]{newman66}
Newman, E., \& Penrose, R.\ 1966,
J. Math. Phys., 7, 863

\bibitem[Noll(1976)]{noll76}
Noll, R.~J.\ 1976,
JOSA, 66, 207

\bibitem[Roodman et~al.(2014)]{rood14}
Roodman, A., Reil, K., \& Davis, C.\ 2014,
Proc.\ SPIE, 9145, 16

\bibitem[Schechter \& Levinson(2011)]{schech11}
Schechter, P., \& Levinson, R.\ 2011,
PASP, 123, 812

\bibitem[Schroeder(1987)]{schroeder87}
Schroeder, D.\ 1987, ``Astronomical Optics'' (Academic Press:San Diego)

\bibitem[Shack \& Thompson(1980)]{shack80}
Shack, R. V., \& Thompson, K.\ 1980,
Proc. SPIE, 251, 146

\bibitem[Stebbins(1996)]{stebbins96}
Stebbins, A.\ 1996, arXiv:9609149

\bibitem[Torres~del Castillo(1992)]{castillo92}
Torres~del Castillo, G.~F.\ 1992,
Rev. Mex. Fis., 38, 19

\bibitem[Trevino et~al.(2013)]{trevino13}
Trevino, J. P. et~al.\ 2013,
Physiol. Opt., 33, 394

\bibitem[Zaldarriaga \& Seljak(1997)]{zald97}
Zaldarriaga, M., \& Seljak, U.\ 1997,
Phys. Rev. D., 55, 1830

\bibitem[Zhao \& Burge(2007)]{zhao07}
Zhao, C., \& Burge, J. H.\ 2007,
Opt. Expr., 15, 18015

\bibitem[Zhao \& Burge(2008)]{zhao08}
Zhao, C., \& Burge, J. H.\ 2008,
Opt. Expr., 16, 6586

\end{thebibliography}
\end{document}